\documentclass[ aps, prb, preprint,superscriptaddress,floatfix, 
longbibliography]{revtex4-2}
\usepackage{amsmath}
\usepackage{amsfonts} 
\usepackage{graphicx}
\usepackage{verbatim}
\usepackage{natbib}
\usepackage[utf8]{inputenc}
\UseRawInputEncoding
\usepackage{booktabs}

\usepackage{multirow}
\usepackage{soul}

\setcitestyle{super}

\begin{document}
\title{Dielectric Screening Inside Carbon Nanotubes}

\author{Georgy Gordeev}
\affiliation{Freie Universit\"at Berlin, Department of Physics, Arnimallee 14, 14195 Berlin}

\author{S{\"o}ren Wasseroth}
\affiliation{Freie Universit\"at Berlin, Department of Physics, Arnimallee 14, 14195 Berlin}

\author{Han Li}
\affiliation{Institute of Nanotechnology, Karlsruhe Institute of Technology, Hermann-von-Helmholtz-Platz 1, 76344 Eggenstein-Leopoldshafen, Germany}

\author{Ado Jorio}
\affiliation{Departamento de F\'isica, Universidade Federal de Minas Gerais, Belo Horizonte, Minas Gerais 30123-970, Brazil}

\author{Benjamin S. Flavel}
\affiliation{Institute of Nanotechnology, Karlsruhe Institute of Technology, Hermann-von-Helmholtz-Platz 1, 76344 Eggenstein-Leopoldshafen, Germany}

\author{Stephanie Reich}
\affiliation{Freie Universit\"at Berlin, Department of Physics, Arnimallee 14, 14195 Berlin}

\begin{abstract}
Dielectric screening plays a vital role for the physical properties in the nanoscale and also alters our ability to detect and characterize nanomaterials by optical techniques. We study the dielectric screening inside of carbon nanotubes and how it changes electromagnetic fields and many-body effects for encapsulated nanostructures. First, we show that the local electric field inside a nanotube is altered by one-dimensional screening with dramatic effects on the effective Raman scattering efficiency of the encapsulated species for metallic walls. The scattering intensity of the inner tube is two orders of magnitude weaker than for the tube in air, which is nicely reproduced by local field calculations. Secondly,  we find that the optical transition energies of the inner nanotubes shift to lower energies compared to a single-walled carbon nanotubes of the same chirality. The shift is higher if the outer tube is metallic than when it is semiconducting. The magnitude of the shift suggests that the excitons of small diameter inner metallic tubes are thermally dissociated at room temperate if the outer tube is also metallic and in essence we observe band-to-band transitions. 

\end{abstract}

\maketitle
\newpage

\begin{figure}
  \centering
  \includegraphics[width=8cm]{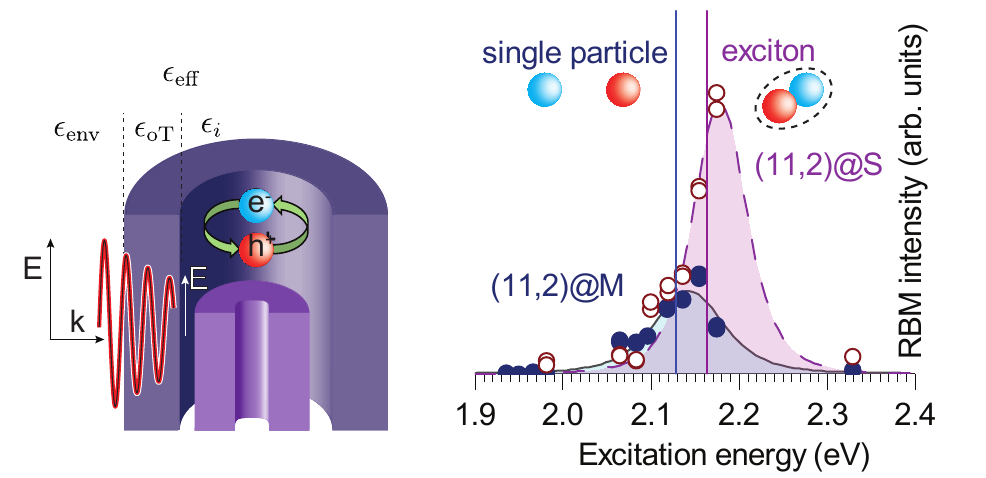}
\end{figure}
\section{Introduction}


Carbon nanotubes (CNTs) can serve as one atom thick container for molecules and  one-dimensional (1D) crystals.\cite{Gallagher1993,Xiang2020,Gaufres2013,Cambr2010,Li2021,Pham2018,Kashtiban2021,Nascimento2021} A CNT container may be exploited as a drug carrier and local sensor.\cite{RaviKiran2020, Huth2018} It also provides a unique environment to tailor and study encapsulated materials. For example, water changes its dielectric behavior and viscosity inside tubes and can adopt new phases.\cite{Kavokine2022,Koga2001} Such water@CNT channels were used for ion transport to potentially model biological systems like transmembrane proteins.\cite{Faucher2021} In another direction, CNTs act as templates to order and align molecules, which led to micron-sized single file $J$ aggregates or molecular chains with huge optical nonlinearities.\cite{Gaufres2013,Cambre2015} Despite extensive studies on filled nanotubes and hybrid one-dimensional (1D) systems, the environment produced by an encapsulating CNT remains mysterious. For instance, 1D chains of dye molecules or carbon atoms inside a nanotube yield record-high Raman cross sections.\cite{Gaufres2013,Tschannen2020} This enhancement may be due to intrinsic effects, molecule-molecule interaction, molecule-wall coupling such as state hybridization, or dielectric effects by the nanotube wall. In case of 1D molecular and carbon chains it is impossible to discriminate between the different effects, because they only exist inside the CNTs and cannot be extracted and studied under ambient conditions. There are some indications for systematic changes of materials inside CNTs. For example, the nanotube walls affect the electromagnetic (EM) field inside the CNT. C$_{60}$@CNT demonstrated different depolarization ratios,\cite{Kuzmany2004} but it remained unclear whether this was related to the depolarization of the EM field or strain. On the other hand, single-walled nanotubes are stable in ambient conditions or can be an inner part of a double-walled CNT (DWCNT).\cite{Gordeev2021} The DWCNTs can potentially serve as ideal probes for the environment produced by a nanotube, since an inner wall nanotube may be easily referenced to an SWCNT\cite{Pesce2010} and will probe the environment produced by the outer wall.\par

The nanotube walls may also alter many-body effects of encapsulated materials, because electron-hole interactions are subjected to the exterior screening.\cite{Thomsen2007} This may fundamentally change how collective states form, for example, for 1D $J$ aggregates inside tubes. It has often been suggested that  nanotube excitons are tuned by interior filling,\cite{Wenseleers2007, Cambr2010} but the inverse effect, where a CNT affects the excitons of encapsulated species has not received much attention. For example, the excitonic series of carbyne are screened by the CNT wall,\cite{Martinati2022} but the intrinsic exciton energies remain unknown. This effect may become particularly noteworthy for metallic outer CNTs, because the metallic species are expected to provide a much denser dielectric environment.\cite{Malic2010} For DWCNTs the screening by an outer tube was predicted to change the excitons of the inner tube to single-particle excitation in small diameter CNTs,\cite{Tomio2012} but experimental evidence has not been reported so far.\par

In this work we study dielectric screening by metallic CNTs using  resonant Raman scattering on DWCNTs. The EM screening by the outer wall reduces the inner tube Raman intensity in metallic@metallic DWCNTs by a factor of $\sim$100, in agreement with a dielectric model of a hollow cylinder in the quasi-static approximation and a dielectric constant of $\epsilon_\mathrm{oT}=10.8$ for the outer metallic wall. We compare the inner tube excitonic transition energies for semiconducting and metallic hosts. The transition energies shift to lower energies compared to SWCNTs, which we analyze in the framework of dielectric screening. The magnitude of the shift in the transitions energies of small inner metallic tubes is compatible with a complete dissociation of the exciton due to dielectric screening.\par

\section{Methods}
To study the dielectric screening effects in the inner walls, we need to sort as-grown DWCNTs into fractions according to the electronic character of the inner and outer wall.\cite{Li2017} We sorted the (inner@outer) DWCNTs using three-step technique into the electronic fractions M@M,M@S, S@M, and M@M, where M indicates metallic and S semiconducting character of the wall. The DWCNTs were filtered first in a gel permeation chromatography column by monitoring Raman intensity. Second, the DWCNTs were re-suspended in toluene and chlorobenzene with PFO-BPy polymer for improved outer-wall separation. At the final step pellets were extracted from solutions by 1 hour centrifugation at up to 10$^6$g and deposited onto silicon substrates.\cite{Li2017} After drying, the samples were used for resonant Raman experiments.

Transition energies and electromagnetic screening were analyzed with resonant Raman spectroscopy of the radial breathing mode (RBM).\cite{Maultzsch2005,Telg2004} Two excitation-tunable lasers were used as excitation sources, for the visible excitation range 570-670nm a Radiant dye laser (DCM, R6G) and a Coherent Ti-Sa laser for near infra-red (700-850nm). The laser was focused onto the sample using a 100x microscope objective (N.A. 0.9) with position and focus optically controlled by a camera. The back-scattered light was filtered by a Horiba triple grating t64000 system to remove the Rayleigh light and dispersed by 600 and 900 grooves/mm gratings. A Peltier cooled charge-coupled device was detecting the Raman signals.  The spectra of 532 nm laser line were acquired with a Horiba Xplora, single-grating spectrometer equipped with a dichroic mirror. The $(n,m)$ chiralities were identified and  RBM shifts were investigated systematically using multi-peak fitting. The concept of laola family facilitated chiral identification, the $(n,m)$ from the same laola group share the parameter $l=2n+m$.  In M@M and M@S samples we found the  $l=2m+n=24$ laola groups -- containing the $(n,m)$ chiralities  (9,6), (10,4), (11,2), and (12,0) -- plus the $l=$27 and 30 laola groups.  A CaF$_2$ single crystal reference spectrum was measured for each laser wavelength. The integrated area of the RBM peaks was divided by the area of CaF$_2$ peak at 320 cm$^{-1}$ in order to account for changes in the sensitivity of optical components in the measurement.\cite{Maultzsch2005}



\section{Results}
\subsection{Theory of dielectric screening in one dimension.}
\begin{figure}
  \centering
  \includegraphics[width=11cm]{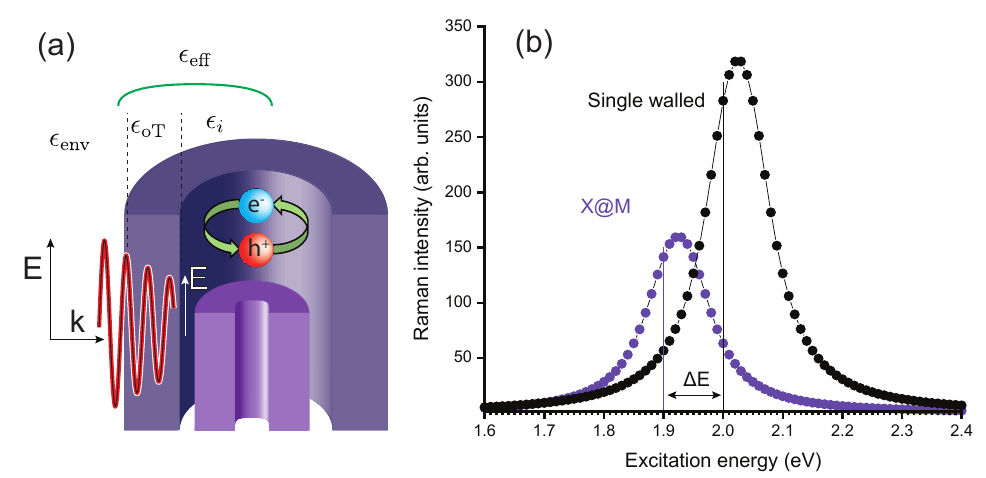}
  \caption{Dielectric screening in DWCNTs. (a) Model of a DWCNT as a cylinder of two dielectric walls. The EM field at the position of the inner tube gets modulated by the outer wall dielectric constant. (b) The dielectric effect manifests in the resonant Raman profiles of the $(n_0,m_0)$  inner wall (blue) compared to corresponding SWCNT (black) computed by Eq. \eqref{EQ:RC1}. The profile of DWCNTs is red-shifted due to exciton screening and the amplitude is reduced by the local field factor compared to the SWCNT.
  }
  \label{FIG:sch}
\end{figure}

We consider a situation where a single-walled carbon nanotubes is embedded in another tube, Fig.~\ref{FIG:sch}a. Neglecting direct tube-tube coupling, the outer tube creates a dielectric environment that changes the optical response of the inner tube in two ways: The outer tube reduces the amplitude of an externally applied electric field for the inner wall and the field orientation so that it is predominantly polarized along the axis.\cite{Thomsen2007}  The change in polarization direction arises from what has been coined the antenna effect,\cite{Thomsen2007} i.e., the fact that a nanoscale cylinder screens an electric field that is polarized perpendicular to its axis.  The change in electric field amplitude is determined by the effective dielectric constant due to the presence of the outer wall, Fig.~\ref{FIG:sch}a. The local field $E_\mathrm{loc}$ inside the outer tube can be derived from classical electrodynamics. We model the outer tube as a hollow cylinder with diameter $d$, wall thickness $t\sim0.35$\,nm, and dielectric constant $\epsilon_\mathrm{oT}$, see Fig.~\ref{FIG:sch}a. The wavelength of light (530-850\,nm) is much larger than the CNT diameter (2\,nm), which indicates the quasistatic regime and field inside of the cylinder can be approximated as\cite{Novotny2012}
\begin{equation}
    E_\mathrm{loc}=E_{o}f_\mathrm{loc}=E_{o}\frac{2\epsilon_\mathrm{env}}{\epsilon_\mathrm{eff}+\epsilon_\mathrm{env}},
    \label{EQ:Loc}
\end{equation}
where $f_\mathrm{loc}$ is the local field factor, $E_0$ external electric field, and $\epsilon_\mathrm{env}$ is a dielectric constant of the environment around the outer tube. The effective dielectric constant $\epsilon_\mathrm{eff}$ arises from the combined dielectric effect of the outer wall itself with $\epsilon_\mathrm{oT}$ and the inner core $\epsilon_i$ ($\epsilon_i=1$ for an empty tube). We calculate it according to Maxwell-Garnet mixing \cite{Holmstrom2010, Sihvola2000, Garnett1904}

\begin{equation}
    \epsilon_\mathrm{eff}=\epsilon_\mathrm{oT}+3 V \epsilon_\mathrm{oT} \frac{\epsilon_i-\epsilon_\mathrm{oT}}{\epsilon_i+2 \epsilon_\mathrm{oT} -V (\epsilon_i-\epsilon_\mathrm{oT})},
    \label{EQ:MG}
\end{equation}
where $V(d)=1-4/(d+t)^2$ is the volume fraction of the empty part of the cylinder. For a CNT with $d = 2$\,nm we find $V = 0.54$. Assuming $\epsilon_\mathrm{i} = 1$ and $\epsilon_\mathrm{o_T} = 10$ we obtain the effective dielectric constant $\epsilon_\mathrm{eff} \approx 5$.

The effective dielectric constant $\epsilon_\mathrm{eff}$ changes the exciton binding energies $E_b$, the electron-electron interaction, and, therefore, the optical transition energy $E_{ii}$ of the inner tube. The exciton binding energy scales as\cite{Walsh2007}

\begin{equation}
    E_{b}(\epsilon)=E_{b}^{\epsilon=1}\epsilon_\mathrm{eff}^{-\alpha}.
    \label{EQ:Eb1}
\end{equation}
For example $E_{b}^{\epsilon=1}\approx 100\,$meV for metallic tubes with $d = 1-2.2$ nm\cite{Malic2010} is the intrinsic binding energy for the unscreened CNT and $\alpha=1.2-1.4$ is a semi-empirical scaling factor\cite{Walsh2007,Perebeinos2004,Araujo2009}. 
The change of the optical transition energy is given by\cite{Ando1997,Walsh2007}
\begin{equation}
    E_{ii}=E_\mathrm{sp} + E_{ee} - E_{eh} = E_\mathrm{sp} + E_\mathrm{BGR}^{\epsilon=1}\epsilon_\mathrm{eff}^{-1} - E_{b}^{\epsilon=1}\epsilon_\mathrm{eff}^{\alpha}
    \label{EQ:Eii}
\end{equation}
The single particle band gap $E_\mathrm{sp}$ is independent of $\epsilon_\mathrm{eff}$. $E_\mathrm{BGR}$ is the electron-electron interaction energy for the unscreened system.\cite{Walsh2007} The electron-electron correlation $E_{ee}$ scales as $ \epsilon_\mathrm{eff}^{-1}$ for small electron wave vectors according to the Coulomb potential.\cite{Kane2004} The electron-hole $E_{eh}$ interaction is more complex and leads back to a hydrogen arom problem in one dimension, where a cutoff potential is typically introduced to converge the ground state  $\sim1/\left|z_0+z \right|$. This yields $E_{eh}=R^*_h/\lambda^2$, where the effective Rydberg radius $ R^*_h $ depends on $\epsilon_\mathrm{eff}$ and $\lambda$ also varies with the potential cutoff $z_0$ as a function of $\epsilon_\mathrm{eff}$. Combining these two factors one gets $E_{eh}\sim \epsilon_{\mathrm{eff}}^{\alpha}$. \cite{Walsh2007} The different scaling of the electron-electron and the electron-hole interaction yields an overall shift of the optical excitation energy as given by Eq.\,\eqref{EQ:Eii}.\par

The optical response of the inner tube is sensitive to the change of the local field and the exciton transition energies caused by the outer tube. In principle, the optical effects can be studied by any optical techniques such as photoluminescence-excitation\cite{Weisman2003}, direct absorption\cite{Streit2018, Pfohl2017}, or resonance Raman spectroscopy\cite{Maultzsch2005,Fantini2004,Gordeev2017}. However, photoluminescence is only present in semiconducting inner tubes and is strongly quenched by the outer wall. Optical absorption is challenging to measure experimentally due to the low cross sections and signal overlap in chiral mixtures. On the other hand, resonant Raman scattering provides sufficient signal for metallic and semiconducting walls up to the single tube level.\cite{Nakar2020} Resonance Raman spectroscopy of the radial breathing modes (RBM) is a key method to follow optical and vibrational changes in carbon tubes.\cite{Maultzsch2005,Fantini2004,Gordeev2017,Thomsen2007}
The phonon energy of the RBM $\hbar\omega_\mathrm{RBM}(d)=c_1/d+c_2$ depends on tube diameter allowing one to distinguish between nanotubes of different size (e.g., inner- and outer-tube). Empirical parameters $c_1$ and $c_2$ depend on to the filling, exterior functionalization, and wall-to-wall interactions. \cite{Telg2004,Fantini2004,Setaro2017,Cambr2010,Gordeev2021}\par

The optical transitions of CNTs are probed via resonance Raman profiles, i.e., the dependence of the scattering intensity on laser excitation.\cite{Telg2004,Fantini2004,Maultzsch2005,Thomsen2007} 
The screening by the outer tube manifests in a shift of the resonant Raman profile that is determined by the optical transition energy $E_{ii}$, Fig.~\ref{FIG:sch}b. The change in field intensity reaching the inner tubes reduces the amplitude of the resonance Raman profile varying with laser excitation energy $E_{l}$  as
\begin{equation}
    I_{R}(E_{l})\propto E_{l}^4{\left[ \frac{M_{R}}{
    (E_{l}-E_{ii}-i\gamma)(E_{l}-\hbar\omega_{ph}-E_{ii}-i\gamma)} \right] }^2
    ,
    \label{EQ:RC1}
\end{equation}
where $M_{R}$ is the combined Raman matrix element. It is given by $M_{R}=M^2_{ex-pt}M_{ex-RBM}$ the product of the exciton-photon $M_{ex-pt}$ and the exciton-phonon matrix elements $M_{ex-RBM}$.\cite{Mueller2016} $M_{ex-pt(in)}$ depends linearly on the local electric field intensity and thus scales with $f^2_{loc}$ which leads to $I_{R}\propto f^4_{loc}$
. The broadening factor $\gamma$ is inversely proportional to the exciton lifetime. The pre-factor $E_{l}^4$ in Eq.~\eqref{EQ:RC1} is eliminated experimentally by calibrating on a Raman reference with a known and constant cross section, see Methods. Figure \ref{FIG:sch}b compares the expected resonant Raman profiles for same ($n_0,m_0$) SWCNT and ($n_0,m_0$)$@$M inner wall of a DWCNT. The energetic positions of the resonance are red shifted due to screening, Eqs.,\eqref{EQ:Eii} and \eqref{EQ:RC1}. The effective matrix element $M_{R}$ for the inner tube is smaller than for the free-standing SWCNT due to the electromagnetic field factor $f_{loc}$, Eqs.~\eqref{EQ:Loc} and \eqref{EQ:RC1}. We expect a smaller RBM intensity from an inner tube of a DWCNT compared to a SWCNT. We now examine the resonance Raman profiles of sorted DWCNTs for signs of the predicted screening effects.\par

\subsection{Inner tube RBM frequency and intensity}

\begin{figure}
  \centering
  \includegraphics[width=9cm]{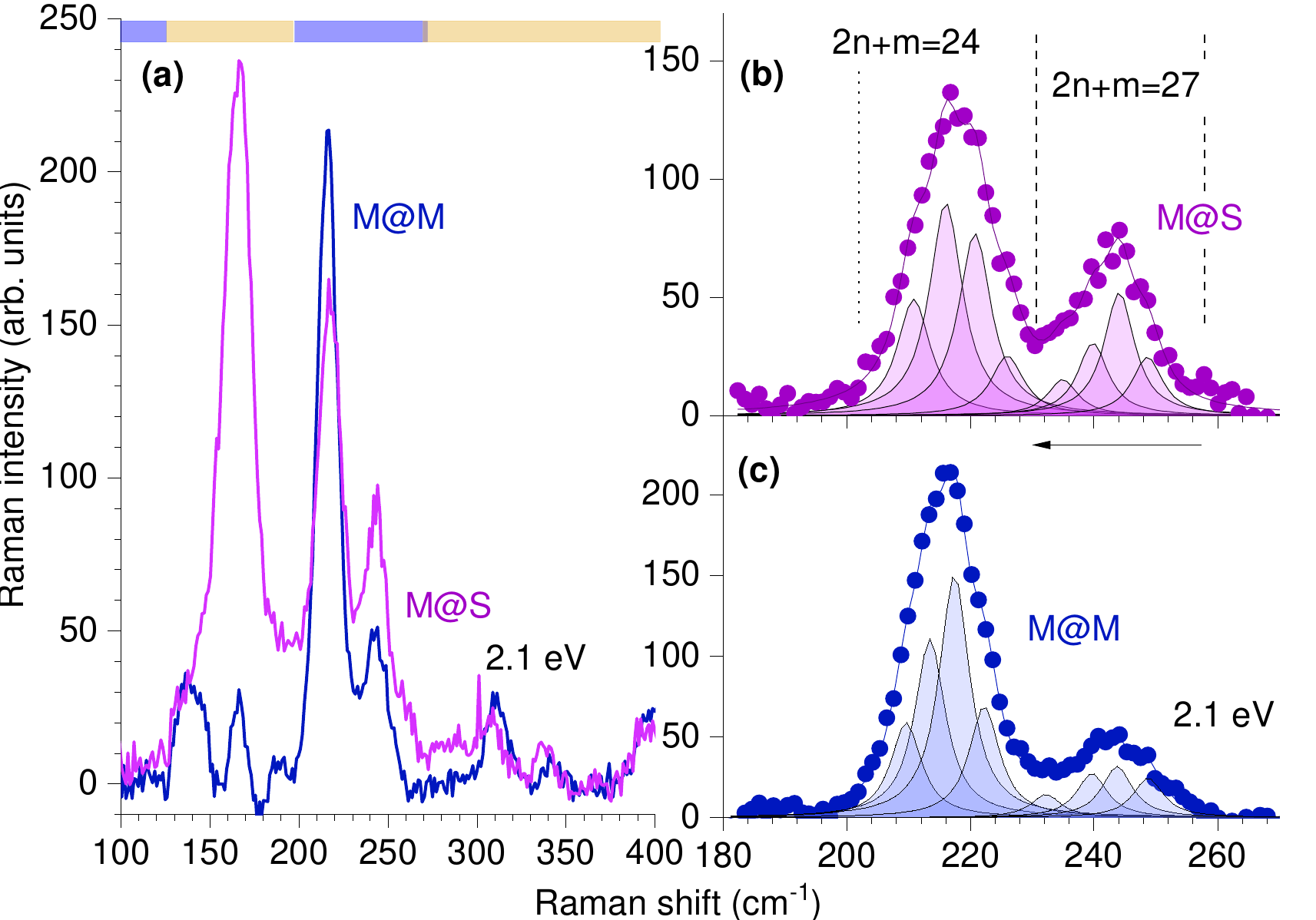}
  \caption{Radial breathing modes in M@M (blue) and M@S (purple) nanotubes excited with a 2.1\,eV laser. (a) RBM spectra in the region between $100-400\,$cm$^{-1}$. The top line roughly divides expected RBMs from metallic (blue) and semiconducting (orange) walls. (b) Inner wall RBM fitting in M@M sample and (c) M@S sample. The vertical lines divide different $2n+m$ laola families. The arrow indicates the increase of the chiral angle within a laola group.}
  \label{FIG:F1}
\end{figure}

 Figure \ref{FIG:F1}a shows the RBM of M@M and M@S DWCNT samples, i.e., all inner tubes are metallic and the outer tubes are metallic in the  M@M but semiconducting in the M@S sample.  We confirm the semiconducting or metallic character of the DWCNT wall from the RBM spectra. We first divide the frequencies into a range of outer $\hbar\omega_\mathrm{RBM}<200\,$cm$^{-1}$ and inner tubes $>200\,$cm$^{-1}$, see labels in Fig.~\ref{FIG:F1}a. We deduce the metallic and semiconducting character of the inner tube from the excitation energy dependence and frequency, since only tubes resonant show measurable intensity.\cite{Thomsen2007,Fantini2004,Maultzsch2005} Thereby we identify the RBM frequency ranges for metallic  (labelled blue) and semiconducting (orange) species, Fig.~\ref{FIG:F1}a.  The M@M and M@S samples contain inner metallic nanotubes with $\hbar\omega_\mathrm{RBM}=200-250\,$cm$^{-1}$. With the laser energy  $E_l=2.1\,$eV in Fig.~\ref{FIG:F1}a we efficiently excite outer semiconducting walls via $E_{33}$; they are strong in the M@S sample and  much weaker in the M@M sample where they appear due to imperfections in the chirality sorting.\cite{Li2017} Resonantly exciting outer metallic walls requires  lower laser energies $E_l<$1.85 eV.\par

The RBM frequencies of the DWCNTs with metallic and mixed walls were only slightly shifted (few cm$^{-1}$ and less) compared to the corresponding SWCNT and no splitting, as it is characteristic for S@S, was observed. This can be explained by the electronic states of inner and outer walls in M@M and M@S being energetically well separated, which reduces moir\'e coupling.\cite{Gordeev2021} We fitted the diameter dependence of the RBM frequencies and obtained a constant $c_1 = 215\,$nm\,$\cdot$cm$^{-1}$ for SW and DWCNTs inner metallic tubes, but a $c_2$ that varied between the three samples [$c_2^{\mathrm{M@M}} = 19.3$\,nm and $c_2^{\mathrm{M@S}} = 20.2$\,nm compared to $c_2^{\mathrm{SWCNTs}}$ = 18\,nm].\cite{Maultzsch2005} The RBM frequency shifts of the inner metallic tubes results from a changed intercept of the RBM diameter dependence, which is associated with environmental effects, as a change in solvent or nanotube bundling.\cite{Maultzsch2005, OConnell2004} \par

\begin{figure}
    \centering
    \includegraphics[width=16cm]{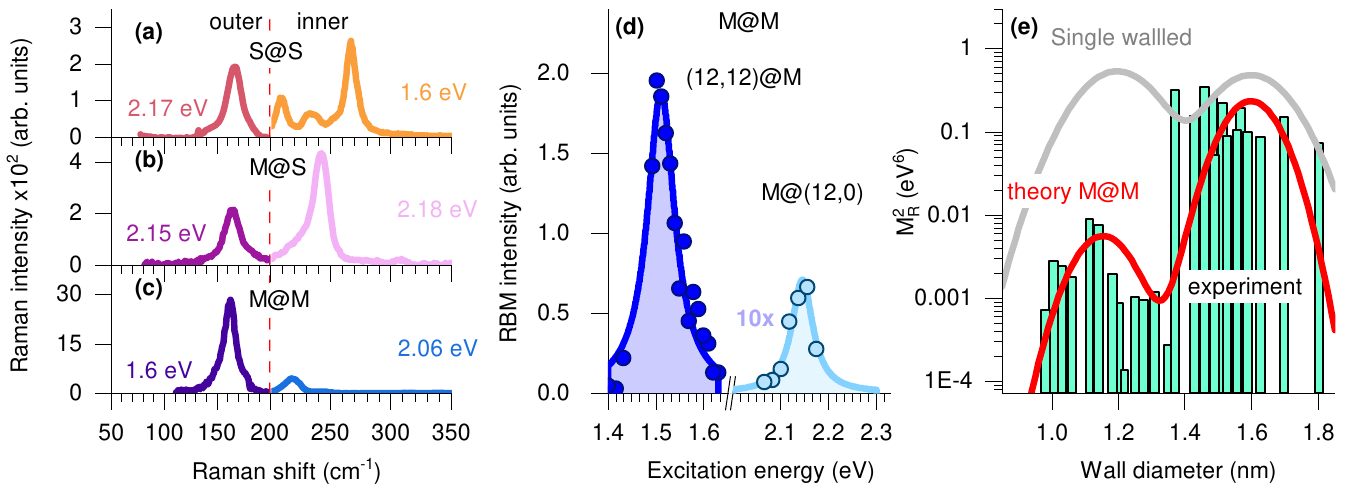}
    \caption{Electromagnetic field screening in the M@M sample. (a,b, and c) Relative Intensities of the RBMs originating from inner and outer walls in S@S, M@S, and, M@M samples, respectively. The inner and outer walls were excited at different energies are separated by a vertical line at 200\,cm$^{-1}$. (d) Resonant Raman profile of the inner (12,0)@M and outer M@(12,12) wall, with RBMs at 249\, and 144\,cm$^{-1}$, respectively. Symbols are experimental data and lines are fits by Eq. \eqref{EQ:RC1}. (e) Green bars: Measured Raman intensities ($M_{R}^2$) as a function of inner wall diameter, estimated from resonance Raman profiles, note logarithmic scale of the y axis. Red line: Calculated Raman intensity $\epsilon_\mathrm{oT}=10$ and $\epsilon_\mathrm{oT}=1$, by Eq. \eqref{EQ:Loc}, Gray line: intrinsic $ M_{R}^2$ without the correction factor. 
    }
    \label{FIG:F3}
  \end{figure}

The EM screening effect by the outer tubes manifests in the measured inner intensities. We first selected spectra with the highest RBM intensities for the inner and outer species in S@S, M@S, and M@M samples, Fig.~\ref{FIG:F3}a-c (the S@M sample had lower sorting purity and will not be considered here). When the outer tubes are semiconducting, top and middle trace, the maximum RBM intensity of the inner tube is comparable or even stronger than that of the outer species. This is expected, because the RBM intensities scale with the inverse of nanotube diameter.\cite{Pesce2010,Machon2005,Thomsen2007} In sharp contrast, the integrated intensity of the inner tube is one order of magnitude smaller with a metallic outer tube in the M@M sample, bottom trace in Fig.~\ref{FIG:F3}c. To study the intensities in detail, we measured the full resonance profiles of the inner and outer walls, see Methods. Figure \ref{FIG:F3}d compares the exemplary Raman profiles of the (12,0) inner ($\hbar\omega_\mathrm{RBM}^i=249\,$cm$^{-1}$, $E_{11}=2.08\,$eV) and the (12,12) outer tube ($\hbar\omega_\mathrm{RBM}^{oT} = 144$ cm$^{-1}$, 1.48\,eV). The difference between the Raman intensities is a striking factors of 30, Fig.~\ref{FIG:F3}d. Similar differences were observed for the other M@M walls as shown in Fig.~\ref{FIG:F3}e for inner and outer tube diameters $0.8-1.8\,$nm.

The experimentally observed variation of Raman intensity is reproduced by Eq. \eqref{EQ:Loc} with dielectric constants $\epsilon_{oT}=9-10$ and $\epsilon_{in}=1$. For fitting our experimental data we represent inner and outer walls populations by two Gaussian like distributions with equal amplitudes, see Supplementary information. The intrinsic diameter-dependent Raman intensity $M^2_R(d)$ for these populations was corrected by the local field factor in Eq. \eqref{EQ:Loc} and plotted in Fig. \ref{FIG:F3}e by the red line, see Supplementary information for more details. Overall, we see good agreement in terms of intensities, with an outer-to-inner intensity ratio of $\sim 36$. The dependence of the $M^2_R$ without local field correction (grey line in  Fig. \ref{FIG:F3}e) unambiguously disagrees with experiment. The experimental dielectric constant $\epsilon_\mathrm{oT}=10$, agrees reasonably well with $\epsilon^{th}_\mathrm{oT}=16$ predicted by \textit{Malic et al}.\cite{Malic2010}


The screening by a typical metallic CNT with $d=1-2\,$nm and $\epsilon_\mathrm{oT}=10$ reduces the amplitude of the electric field within the tube by $f_{loc}\approx 0.36$ compared to the far field. This has important consequences when CNTs are used as containers or reactors and the encapsulated species are monitored by optical methods. The linear optical response of encapsulated molecules will drop by a factor of $f_{loc}^2 \approx 0.13$; in the case of non-linear techniques such as Raman scattering, the reduction amounts to two orders of magnitude ($1.6\cdot 10^{-2}$), Fig. \ref{FIG:F3}e. The sharp drop in intensity, may be the reason why carbyne chains so far were only observed in individual semiconducting DWCNTs.\cite{Heeg2018, Heeg2018a} An in-situ monitoring of chemical reactions, e.g., in the fusion of molecules to graphene ribbons and inner tubes,\cite{KUZMANY2021221,Cadena2022} will strongly favor reactions inside semiconducting containers, although metallic tubes might actually be superior for that task. On the other hand, an exciting concept that follows from Eq. \eqref{EQ:Loc} is active screening when the laser frequency matches the optical transition of the outer tube. The dielectric function then follows the Lorentz resonance and creates an additional active dielectric screening. Matching an inner and outer tube resonance in DWCNTs is best achieved in  S@S or mixed electronic type samples. Realizing active outer screening on individual DWCNTs with overlapping optical resonances would be of great interest for future work.

\subsection{Exciton screening and optical transition energies shift.}

\begin{figure}
    \centering
    \includegraphics[width=8cm]{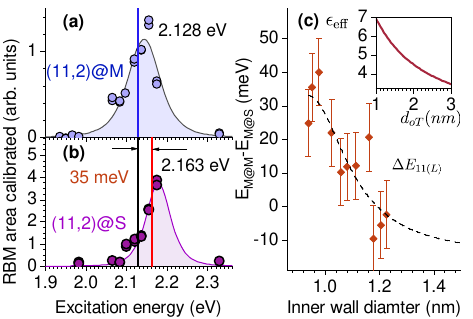}
    \caption{Exciton screening in the S@M and M@M samples studied by resonant Raman scattering. (a) Resonant Raman profile of 
 (11,2)@M compared to (b) (11,2)@S. Symbols represent experimental data and lines fits by Eq.~\eqref{EQ:RC1}. The positions of the transition energies are marked by vertical lines. (c) Energy shift of the transitions measured in identical inner tubes in the M@M and M@S samples, listed in Table~\ref{TAB:T1}. The inset shows the diameter dependence $\epsilon_\mathrm{eff}$, versus outer wall diameter for $\epsilon_\mathrm{{oT},{i}}=10,1$ %
 }
    \label{FIG:F2}
  \end{figure}

We determine the transition energies of inner CNTs from the resonant Raman profiles. We start with the peak at 244\,cm$^{-1}$ in Fig.~\ref{FIG:F1}a, belonging to the RBM (11,2)@S. The integrated area of the (11,2)@S RBM is plotted as a function of excitation energy in Fig.~\ref{FIG:F2}b. The intensity increases when the laser approaches the $E_{11(L)}$  transition energy of the nanotube. We quantify this energy by fitting the (11,2)@S Raman profile by Eq. \eqref{EQ:RC1} and find the transition energy $E_{11(L)}$=2.16 eV, marked by a vertical red line. The transition energy in the DWCNTs is comparable to the (11,2) SWCNTs ($2.12\,$eV). In single-walled CNTs the $E_{11(L)}$ may depend on many factors, surfactant type and filling, compared to DWCNTs where the exclusive environment is the outer wall. Next, we investigate the transition energy for a metallic outer wall.

 The metallic outer wall induces a larger transition energy of the inner wall exciton, compared to a semiconducting outer wall. The resonance Raman profiles for (11,2)@S and (11,2)@M are plotted in Figures \ref{FIG:F2}b and c, respectively. In the (11,2)@M sample, we find $E_{11(L)}^{(11,2)@M}= 2.163$ eV, which is smaller by 35 meV compared to $E_{11(L)}^{(11,2)@S}$. Since the Raman frequencies of the inner walls are identical, both at $\sim$244 cm$^{-1}$, moir\'e coupling can be neglected \cite{Gordeev2021} and the dominant origin of the energy shift must be dielectric screening, as expected from Eq. \eqref{EQ:Eii}.\par 

The red shift in transition energy is stronger for small-diameter inner nanotubes. We plot the shift of transition difference between the M@S and M@M samples in Fig. \ref{FIG:F2}c. The magnitude of the shift decreases from $\sim$40 meV  for d = 1\,nm to nearly zero d = 1.2\,nm. The shift scales with the effective dielectric constant $\epsilon_\mathrm{eff}$, plotted as a function of inner wall diameter for $\epsilon_\mathrm{{oT},{i}}=10,1$ in the inset of Figure \ref{FIG:F2}c. The transition energies shifts in all other investigated chiralities are listed in Table \ref{TAB:T1}. The scattering of the symbols in the order of 10\,meV is mainly due to the weak moir\'e effects, which as well induce a slight negative shift in some cases. \par


\begin{table}[h]
    \centering
    \caption{Summary of RBM frequencies $\hbar\omega_{ix}$ and transition energies $E_{11(L)}^{ix}$ extracted from fitting resonance Raman profiles, where $ix=$ SW (single-walled), M@M, and M@S. $\Delta E_{11(L)}^{M@X}=E_{11}^{M@S}-E_{11}^{M@M}$.}
    \begin{tabular}{lll lll lll}
    
        \toprule 
 $2n+m$ & $(n,m)$ & $\hbar \omega_{SW}$ & $\hbar \omega_{M@M}$ & $\hbar \omega_{M@S}$ & $E_{11(L)}^{SW}$ & $E_{11(L)}^{M@M}$ & $E_{11(L)}^{M@S}$  & $\Delta E_{11(L)}^{M@X}$  \\
  &  & cm$^{-1}$ & cm$^{-1}$ & cm$^{-1}$ & eV & eV & eV  & meV  \\
  \midrule
24 & (9,6)     & 228       & 232.0   & 231.7  & 2.25    & 2.134   & 2.156 & 22             \\
  & (10,4)    & 238       & 238.8   & 238.9  & 2.24    & 2.132   & 2.172 & 40             \\
  & (11,2)    & 244       & 244.2   & 244.5  & 2.22    & 2.128   & 2.163 & 35             \\
  & (12,0)    & 247       & 249.4   & 249.6  & 2.21    & 2.128   & 2.152 & 25             \\
27 & (10,7)    & 203       & 207     & 210    & 2.04    & 2.050   & 2.070 & 21             \\
  & (11,5)    & 211       & 210     & 213    & 2.08    & 2.050   & 2.062 & 12             \\
  & (12,3)    & 217       & 217     & 217.9 & 2.07    & 2.037   & 2.049 & 12             \\
  & (13,1)    & 221       & 222     & 222.4  & 2.07    & 2.024   & 2.034 & 10             \\
30 & (13,4)    & 193.5     & 195.9   & 197.6  & 1.93    & 1.902   & 1.899 & -2             \\
  & (14,2)    & 196.3     & 199.8   & 201.9  & 1.92    & 1.901   & 1.896 & -6             \\
  & (15,0)    & 200.4     & 204.4   & 206.0  & 1.88    & 1.904   & 1.894 & -10   \\
\bottomrule
    \end{tabular}
    \label{TAB:T1}
\end{table}



Excitons in thinner M@M inner walls can be reduced to single particles by the extreme outer wall screening. The thermal dissociation of the excitons occurs when the exciton binding energy becomes comparable to the thermal energy at room temperature $k_B T_{293} = 25\,$meV.  With $\mathrm{\epsilon_{oT}}= 10$ we obtain $\mathrm{\epsilon_{eff}}= 4.5$ and a reduction in binding energy from 114 to 15\,meV. This yields a $E_{11(L)}$ red-shift of 19\,meV, given by Eqs. \eqref{EQ:Eb1} and \eqref{EQ:Eii}. Such or greater red shifts are indeed observed for inner CNTs with $d<1.03$\,nm in our sample, Fig.~\ref{FIG:F3} and Table \ref{TAB:T1}, indicating that we observe single particle band gaps in the inner M@M walls. 

Many fundamental properties of CNTs are governed by excitons, including absorption\cite{Malic2010}, emission\cite{Maultzsch2005b}, and Raman scattering.\cite{Gordeev2019,Gordeev2017} It would be extremely interesting to redo such experiments on screened excitons as available in the inner walls of M@M DWCNTs. For example, we expect asymmetric absorption peaks, identical energies in one- and two-photon luminescence excitation spectroscopy, and a change in the relative intensity of the incoming and outgoing G mode Raman resonances.\cite{Maultzsch2005b,Tomio2012,Gordeev2017, Gordeev2019} The fragility of the excitonic states in partly or fully metallic DWCNTs makes them unlikely candidates for preparing exciton condensates and exciton insulators, where a better choice would be fully semiconducting species with strong moir\'e effects.\cite{Gordeev2021}  On the other hand, the metallic walls would be better reactors due to the low-energy electronic bands, however much lower optical signals will require enriched metallic nanotube samples for detection of reaction products.  

The dielectric effects may be used to control materials encapsulated inside carbon nanotubes via screening. The optical transition energies of the molecules and carbon chains are ruled by excitonic effects.\cite{Pedersen2017} Such effects would manifest in an optical energy shift when confined inside metallic nanotubes compared to semiconducting ones. For example, in linear carbon chains, we would expect a deviation from the linear behavior between the Raman mode frequency and the transition energy.\cite{Shi2017} Up to now, individual single carbon chains have only been reported in semiconducting CNT containers.\cite{Tschannen2020,Tschannen2021} This is likely related to their localization method, where first the lateral Raman maps are analyzed for the highest Raman signal. As we showed, the Raman signals inside the metallic shells are much smaller, therefore, improved localization methods are required to target excitonic effects in one-dimensional chains.

\section{Conclusions}

Dielectric screening plays an important role in DWCNT; its effects are two-fold as it modulates many-body effects and alternates the electric field inside the outer tube. Many-body effects manifest in the energetic position of the excitons. We measured inner walls exciton energies by means of resonant Raman spectroscopy for semiconducting and metallic outer walls. In metallic outer walls we found an additional red-shift by up to 40\,meV, compared to semiconducting outer walls. The optical resonances of inner metallic walls most likely originate from band-to-band excitations, because the excitons dissociate thermally. The electric field is also strongly altered by the electronic type of the outer wall. The metallic walls act as a dense dielectric shield blocking a substantial fraction of the electromagnetic field. That manifests in up to 30 times weaker Raman signals of the inner metallic walls compared to the outer metallic walls. These results open interesting prospects for dielectric cloaking and active dielectric screening. We believe that in all types of one-dimensional heterostructures one will find strong dielectric effects altering many-body interactions and electromagnetic fields.

\section{Acknowledgments}
G.G. and S.R. acknowledge the Focus Area NanoScale of Freie
Universitaet Berlin and the supraFAB Research Center. S.R. acknowledges support by the
Deutsche Forschungsgemeinschaft under Grant SPP 2244
and the European Research Council ERC under Grant DarkSERS. BSF acknowledges support from the DFG under grant numbers: FL 834/5-1, FL 834/7-1, FL 834/12-1, FL834/13-1, FL 834/9-1

\section{References}
\bibliography{bibliography}
\end{document}


\title{Supplementary information: Dielectric effects in one-dimensional heterostructures}

\author{Georgy Gordeev}
\affiliation{Freie Universit\"at Berlin, Department of Physics, Arnimallee 14, 14195 Berlin}

\author{S{\"o}ren Wasserroth}
\affiliation{Freie Universit\"at Berlin, Department of Physics, Arnimallee 14, 14195 Berlin}

\author{Han Li}
\affiliation{Institute of Nanotechnology, Karlsruhe Institute of Technology, Hermann-von-Helmholtz-Platz 1, 76344 Eggenstein-Leopoldshafen, Germany}

\author{Ado Jorio}
\affiliation{Departamento de Física, Universidade Federal de Minas Gerais, Belo Horizonte, Minas Gerais 30123-970, Brazil}

\author{Benjamin S. Flavel}
\affiliation{Institute of Nanotechnology, Karlsruhe Institute of Technology, Hermann-von-Helmholtz-Platz 1, 76344 Eggenstein-Leopoldshafen, Germany}

\author{Stephanie Reich}
\affiliation{Freie Universit\"at Berlin, Department of Physics, Arnimallee 14, 14195 Berlin}

\maketitle
\tableofcontents

\newpage

\section{Experimental Methods}
To study the dielectric screening effects in the inner walls, we need to sort as-grown DWCNTs into fractions according to the electronic character of the inner and outer wall.\cite{Li2017} We sorted the (inner@outer) DWCNTs using a three-step technique into the electronic fractions M@M,M@S, S@M, and M@M, where M indicates metallic and S semiconducting character of the wall. The DWCNTs were filtered first in a gel permeation chromatography column by monitoring the Raman intensity. Second, the DWCNTs were re-suspended in toluene and chlorobenzene with PFO-BPy polymer for improved outer-wall separation. At the final step pellets were extracted from solutions by 1 hour centrifuging at up to 10$^6$g and deposited onto silicon substrates.\cite{Li2017} After drying, the samples were used for resonant Raman experiments. The Raman maps from four electronic fraction were used to confirm the purity of the samples in ref. \citet{Li2017} The highest purity is found in the S@S,M@M, and M@S samples, whereas in the S@M sample we observed undesirable signal from semiconducting outer walls originating from the $E_{33}$ transition (1.6 nm, 2.3 eV). Further the average diameter of the inner walls was higher compared to the S@S sample with an average of 1 nm.

Transition energies and electromagnetic screening were analyzed with resonant Raman spectroscopy of the radial breathing mode (RBM).\cite{Maultzsch2005,Telg2004} Two excitation-tunable lasers were used as excitation sources, for the visible excitation range 570-670nm a Radiant dye laser (DCM, R6G) and a Coherent Ti-Sa laser for near infra-red excitation (700-850nm). The laser was focused onto the sample using a 100x microscope objective (N.A. 0.9) with position and focus optically controlled by a camera. The back-scattered light was filtered by a Horiba triple grating t64000 system to remove the Rayleigh light and dispersed by 600 and 900 grooves/mm gratings. A Peltier cooled charge-coupled device was detecting the Raman signals.  The spectra of 532 nm laser line were acquired with a Horiba Xplora, single-grating spectrometer equipped with a dichroic mirror. The $(n,m)$ chiralities were identified and  RBM shifts were investigated systematically using multi-peak fitting. The concept of laola family facilitated chiral identification, the $(n,m)$ from the same laola group share the parameter $l=2n+m$.  In M@M and M@S samples we found the  $l=2m+n=24$ laola groups -- containing the $(n,m)$ chiralities  (9,6), (10,4), (11,2), and (12,0) -- plus the $l=$27 and 30 laola groups.  A CaF$_2$ single crystal reference spectrum was measured for each laser wavelength. The integrated area of the RBM peaks was divided by the area of CaF$_2$ peak at 320 cm$^{-1}$ in order to account for changes in the sensitivity of optical components in the measurement.\cite{Maultzsch2005}

\section{Variation of Raman intensity in SWCNTs and in the M@M sample}
In this section we explain intensities variation in the single walled carbon nanotubes and how local factor can be implemeted. The population is modelled by a standard Gaussian function:
\begin{equation}
    I_{i,o}(d)=Ae^{-\frac{(d-b)^2}{2c^2}}
    \label{SEQ:Gaus}
\end{equation}

We also need to account for chiral variation of matrix elements\cite{Jiang2007a}. For SWCNTs  can use empirical formulas proposed by \textit{Pesce et al}.\cite{Pesce2010}
\begin{equation}
    M_R(d)=\left(M_A+\frac{M_B}{d}+\frac{M_C \cos(3 \Theta) }{d^2}\right)^2,
    \label{SEQ:Mii}
\end{equation}
where $M_i$ with $(i = A,B,C)$ are constants ($M_A=1.68 eV, M_B = 0.52 nm\cdot eV, M_C = nm^2\cdot eV$) deduced by correlating transmission electron microscopy (TEM) and resonance Raman scattering,\cite{Pesce2010} $d$ is the CNT diameter and $\Theta$ its chiral angle.\cite{Thomsen2007}

The full profile of Raman matrix elements can be described as 
\begin{equation}
    M_R(d)=M_R(d)\cdot( f_{loc}^4(\epsilon_{eff}) I_i(d) + I_o(d)),
    \label{EQS:Population}
\end{equation}

Transition energies of the CNTs can be calculated using following empirical formula\cite{Araujo2009}
\begin{equation}
    E_{ii}(p,d)-\beta_p cos3\theta/d^2 = a\frac{p}{d}\left[1+b \cdot log\frac{cd}{p}\right],
    \label{SEQ:Eii}
\end{equation}
\noindent where p = (1,2,3,4,5) for ($E_{11}^S, E_{22}^S, E_{11}^M, E_{33}^S, E_{44}^S$), a = 1.049 eV$\cdot$nm, b = 0.456, and c = 0.812 nm$^{-1}$. The $\beta_p$ parameter depends on the transition number and CNT type. The CNT is of type 1 when $(2n+m)mod3 = 1$ and of type 2 if $(2n+m)mod3 = 2$. The $\beta_p$ equals (-0.07,0.05), (0.19,-0.14), (-0.19), (0.42,0.42), and (0.4,0.4) for  p = 1,2,3,4, and 5 respectively. \par
\begin{figure}
  \centering
  \includegraphics[width=14cm]{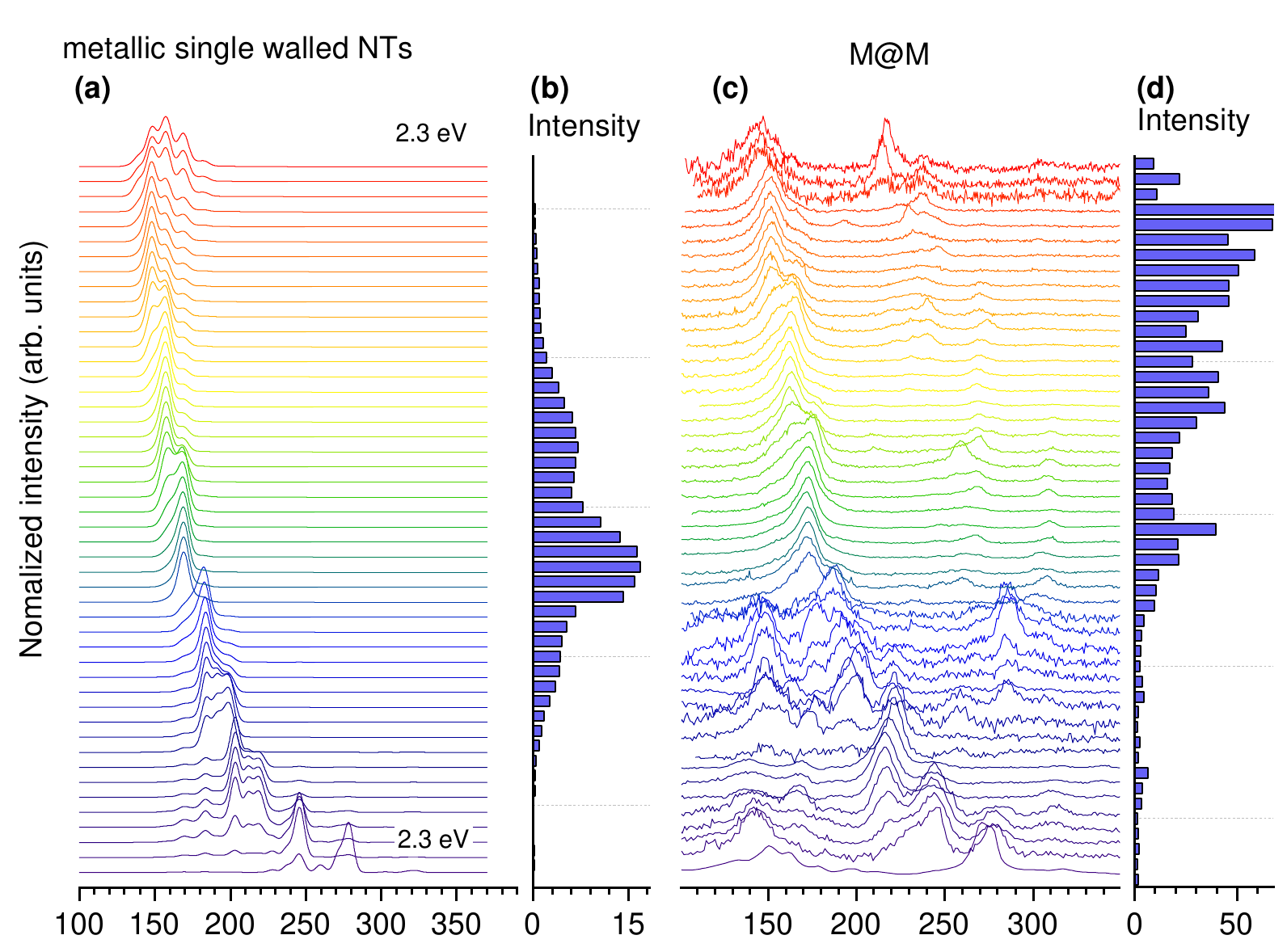}
  \caption{Resonant Raman spectra of nanotubes excited between of \SI{2.3}{\eV} (blue) and \SI{1.55}{\eV} (red), the intermediate colors are mixed in rainbow-like pattern. The panel (a) represents the single-walled CNTs, generated for the same populations as M@M walls using eqs. \eqref{SEQ:RBM_M},\eqref{SEQ:Eii}, \eqref{SEQ:Mii}. \cite{Araujo2009,Pesce2010}. The spectra are normalized and offset for clarity, (b) the normalization factors for each laser wavelength. (c) Experimental spectra of M@M sample, with c) histograms of the maximum Raman intensity.   }
  \label{FIG:exp}
\end{figure}
In Figure \ref{FIG:exp}a an b we show the waterfall plots for SWCNT and M@M sample, the RBMs belonging to different chiralities move inside and outside the resonance conditions, depending on the laser energy. The SWCNT spectra were obtained  using known intensity and transition energies behaviors\cite{Araujo2009,Pesce2010} and the M@M are the experimental. The spectra are normalized to one for better visibility and the normalization factors are shown in the right panels. In the M@M sample we find only metallic walls, compared to the SWCNT sample, shown in Figure \ref{FIG:exp} b,a. The M@M and S@S samples are the cleanest in the separation process.\cite{Li2017} Interestingly, the intensities distributions are inverse. In the single walled CNTs a highest intensity is found for smaller diameter tubes\cite{Pesce2010}, whereas in the M@M sample we find largest intensity for the large diameter walls, red region in Figure \ref{FIG:exp}b. This inversion already indicates dielectric effects. When analyzing the M@S sample we find some smaller semiconducting inner walls, see Figure \ref{FIG:exp}c and overall intensity behavior is more erratic. This is caused by S@S impurities and also different $E_{ii}$ transitions at play. Hence, we concentrate only on the M@M sample for studying the EM screening. The M@M an ideal sample, since both inner and outer walls are excited at the same transition. However, first, we need to identify individual (n,m) chiralities from the groups the RBM peaks in Figure \ref{FIG:exp}.
\section{RBM frequency diameter dependence}
\begin{figure}
    \centering
    \includegraphics[width=9cm]{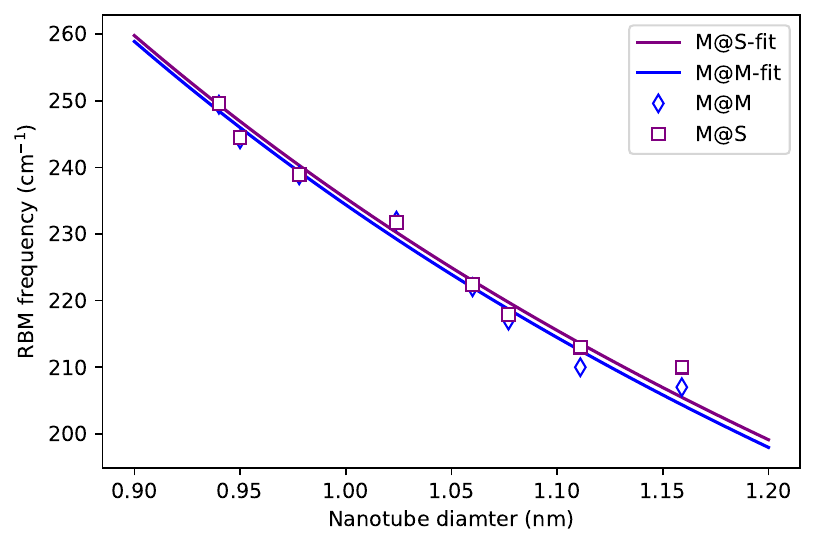}
    \caption{Fit of the experimental RBM frequencies with Eq. \eqref{SEQ:RBM_2}}
    \label{SIG:RBM_fit}
\end{figure}

The dependence of the RBM frequency depends on the effective mass of the CNT cylinder, traditionally the expression was used:\cite{Maultzsch2005}
\begin{equation}
    \omega_{RMB}(d) = c_1/d +c_2,
    \label{SEQ:RBM_M}
\end{equation}
where $c_1$ and $c_2$ are constants determinned from an experiment. Later, a more physical expression was proposed:\cite{Araujo2008}
\begin{equation}
    \omega_{RMB}(d) = 227\sqrt{(\frac{1}{d^2}+\frac{6(1-v^2)}{Eh} \frac{K}{s_0^2} )},
    \label{SEQ:RBM}
\end{equation}
where $\frac{6(1-v^2)}{Eh} = 26.3$ meV is intrinsically determined constant and $\frac{K}{s_0^2}$ incorporates interaction forces with the CNTs environment. We combine these two constants into $c_2$ to simplify the expression:
\begin{equation}
    \omega_{RMB}(d) = 227\sqrt{(\frac{1}{d^2}+c_a )},
    \label{SEQ:RBM_2}
\end{equation}
where $c_a =\frac{6(1-v^2)}{Eh} \frac{K}{s_0^2}  $. Figure \ref{SIG:RBM_fit} shows the fit by Eq. \eqref{SEQ:RBM_2}. We obtain $c^{ms}_a$ = 2.85, $c^{mm}_a$ = 2.5, both slightly higher than $c^{swcnts}_a$ 2.2.\cite{Araujo2008}

\begin{table}[h]
    \centering
    \caption{Summary of RBM frequencies $\hbar\omega_{M@S}$ and transition energies $E_{11(L)}^{M@X}$, where $X=M$ for metallic inner wall and $X=S$ for semiconducting inner wall. $\Delta E_{11(L)}^{M@X}=E_{11}^{SW}-E_{11}^{DW}$. $\Delta E_{11(L)}^{M@X}$ is $E_{11(L)}^{M@M}-E_{11(L)}^{M@S}$ .The data for SWCNTs is a combination of empirical and calculated values from \textit{Maultzsch et al.}\cite{Maultzsch2005}, see Methods. In Eq. \eqref{SEQ:RBM_M} $c^{M@M}_2 = 19.3nm$, $c^{M@S}_2 = 20.2nm$, and $c_1= 215 cm^{-1}\cdot nm$}
    \begin{tabular}{lllllllllll}
    
        \toprule
        chirality & $\omega_{SW}$ & $E_{SW}$ & $\hbar\omega_{M@M}$ & $E_{11(L)}^{M@M}$ & $\hbar\omega_{M@S}$ & $E_{11(L)}^{M@S}$ & $\Delta E^{X@M}$ & $\Delta E_{11(L)}^{M@M}$ & $\Delta E_{11(L)}^{M@S}$ & $\Delta\hbar\omega$ \\ 
        $(n,m)$ & cm$^{-1}$ & eV & cm$^{-1}$ & eV & cm$^{-1}$ & eV & meV & meV & meV & cm$^{-1}$ \\ \midrule
        $l$=24 &  &  &  &  &  &  &  &  &  &  \\
        (9,6) & 228 & 2.25 & 232.0 & 2.134 & 231.7  & 2.156 & 22 & -116 & -94 & 3.7   \\
        (10,4)    & 238 & 2.24 & 238.8 & 2.132 & 238.9  & 2.172 & 40 & -108 & -68 & 0.9   \\
        (11,2)    & 244 & 2.22 & 244.2 & 2.128 & 244.5  & 2.163 & 35 & -92  & -57 & 0.5   \\
        (12,0)    & 247 & 2.21 & 249.4 & 2.128 & 249.6  & 2.152 & 25 & -82  & -58 & 2.6   \\ 
        $l$=27 &  &  &  &  &  &  &  &  &  &  \\
        (10,7)    & 203 & 2.04 & 207   & 2.050 & 210    & 2.070 & 21 & 10   & 30  & 7     \\
        (11,5)    & 211 & 2.08 & 210   & 2.050 & 213    & 2.062 & 12 & -30  & -18 & 2     \\
        (12,3)    & 217 & 2.07 & 217   & 2.037 & 217.9 & 2.049 & 12 & -33  & -21 & 0.9  \\
        (13,1)    & 221 & 2.07 & 222   & 2.024 & 222.4  & 2.034 & 10 & -46  & -36 & 1.4\\
        $l$=30 &  &  &  &  &  &  &  &  &  &  \\
        (13,4) & 193.5 & 1.93 & 195.9 & 1.902 & 197.6 & 1.899 & -2  & -28 & -31 & 4.1  \\
(14,2) & 196.3 & 1.92 & 199.8 & 1.901 & 201.9 & 1.896 & -6  & -19 & -24 & 5.6  \\
(15,0) & 200.4 & 1.88 & 204.4 & 1.904 & 206.0 & 1.894 & -10 & 24  & 14  & 5.6  \\
\bottomrule
    \end{tabular}
    \label{TAB:T1}
\end{table}

\section{Maxwell-Garnett mixing}

\begin{figure}
  \centering
  \includegraphics[width=12cm]{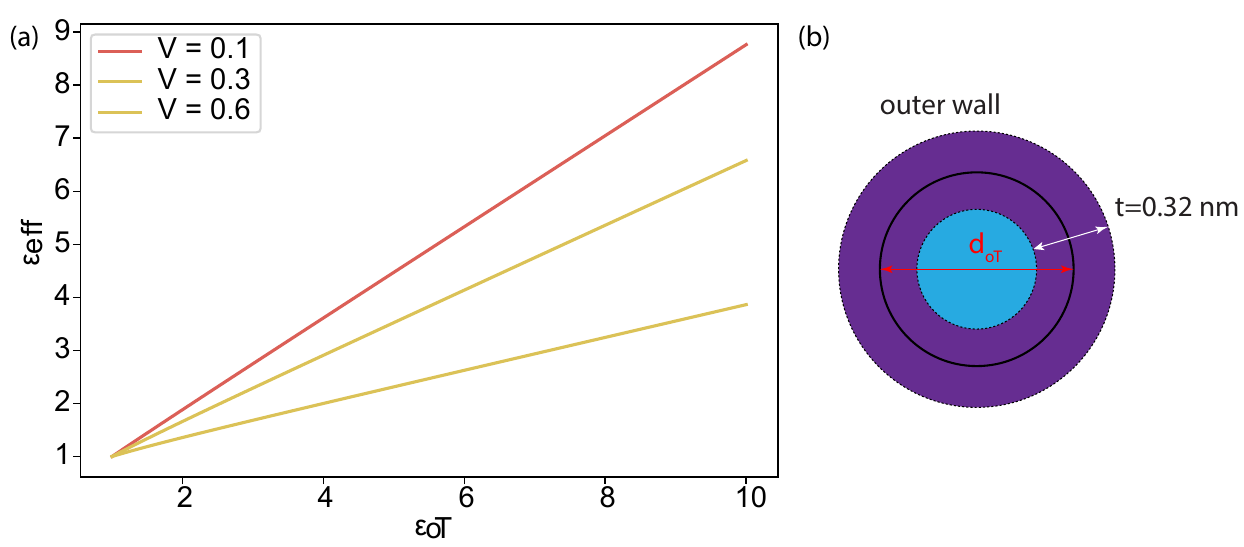}
  \caption{ (a) Dependence of $\epsilon_\mathrm{eff}$ on the outer wall dielectric constant $ \epsilon_\mathrm{oT} $, for different volume fractions calculated by Eq. \eqref{EQ:MGQD} and (b) scheme of volume fraction calculation.}
  \label{FIG:MG}
\end{figure}

The formula Maxwell-Garnet mixing used \cite{Holmstrom2010}
\begin{equation}
    \epsilon_\mathrm{eff}=\epsilon_{oT}+3 V \epsilon_\mathrm{oT} \frac{\epsilon_i-\epsilon_\mathrm{oT}}{\epsilon_i+2 \epsilon_\mathrm{oT} -V (\epsilon_i-\epsilon_\mathrm{oT})}.
    \label{EQ:MGQD}
\end{equation}
\noindent where $V(d)=1-4/(d+t)^2$ is the volume fraction of the empty part of the cylinder. Eq \eqref{EQ:MGQD}is plotted in Figure \ref{FIG:MG} for various  $V$ parameters. \par

The volume fraction $V$ refers to the ratio of the outer nanotube empty $V_{emp}$ part to the outer nanotube 'full' volume $V_{full}$. The scheme is shown in figure \ref{FIG:MG}, with empty part coloured in blue and filled part coloured in purple. The empty part is $\pi (r_\mathrm{oT}-t/2)^2 h$, whereas the full part is $\pi (r_\mathrm{oT}+t/2)^2 h$, with $h$ being the nominal length of the nanotube and $r$ its radius. The overall volume of the unfilled to the total volume is therefore:

\begin{equation}
    V_{emp} = \frac{V_{unfilled}}{V_{total}}=\frac{(r_\mathrm{oT}-t/2)^2}{(r_\mathrm{oT}+t/2)^2}= \frac{(d_\mathrm{oT}-t)^2}{(d_\mathrm{oT}+t)^2}.    
    \label{EQ:V}
\end{equation}

\bibliography{bibliography}